\documentclass[runningheads,a4paper]{llncs}
\pdfoutput=1 
\usepackage{amssymb,amsmath}
\usepackage{graphicx}
\usepackage{enumerate}
\usepackage{wrapfig}
\usepackage[font=small, labelfont=bf]{caption}
\usepackage{subfig}

\newcommand{\ttree}[1]{\ensuremath{\langle #1 \rangle}}
\newcommand{\recsplit}{\emph{rec-split}}
\newcommand{\recimproved}{\emph{rec-split-improved}}
\newcommand{\recbb}{\emph{rec-split-bb}}
\newcommand{\hiersort}{\emph{hierarchy sort}}

\begin{document}

\title{Drawing Binary Tanglegrams: \\[.5ex] \large An Experimental Evaluation}

\titlerunning{Drawing Binary Tanglegrams: An Experimental Evaluation}

\author{
Martin N\"ollenburg\inst{1}\thanks{Supported by grant WO 758/4-3 of
  the German Research Foundation (DFG).} \and Danny Holten\inst{2} \and Markus V\"olker\inst{1} \and Alexander Wolff\inst{2}
}

\institute{%
  Fakult{\"a}t f{\"u}r Informatik, Universit\"{a}t Karlsruhe, Germany. \\
  $\{$ \email{noellenburg}, \email{mvoelker} $\}$ \email{@iti.uka.de}
  \and
  Faculteit Wiskunde en Informatica, TU~Eindhoven, The~Netherlands. \\
  \email{d.h.r.holten@tue.nl},
  \email{http://www.win.tue.nl/\~{}awolff}
}

\maketitle

\begin{abstract}
  A \emph{binary tanglegram} is a pair \ttree{S,T} of binary trees
  whose leaf sets are in one-to-one correspondence; matching leaves
  are connected by inter-tree edges.  For applications, for example in
  phylogenetics or software engineering, it is required that the
  individual trees are drawn crossing-free. A natural optimization
  problem, denoted \emph{tanglegram layout problem}, is thus to
  minimize the number of crossings between inter-tree edges.

  The tanglegram layout problem is NP-hard and is currently considered
  both in application domains and theory.  In this paper we present an
  experimental comparison of a recursive algorithm of Buchin et
  al.~\cite{bbbnow-dbtha-08}, our variant of their algorithm, the
  algorithm hierarchy sort of Holten and van Wijk~\cite{hw-vchod-08},
  and an integer quadratic program that yields optimal solutions.
\end{abstract}

\section{Introduction}

\begin{figure}[tb]
  \subfloat[arbitrary layout\label{sfg:permuted}]%
  {\includegraphics[width=.48\textwidth]{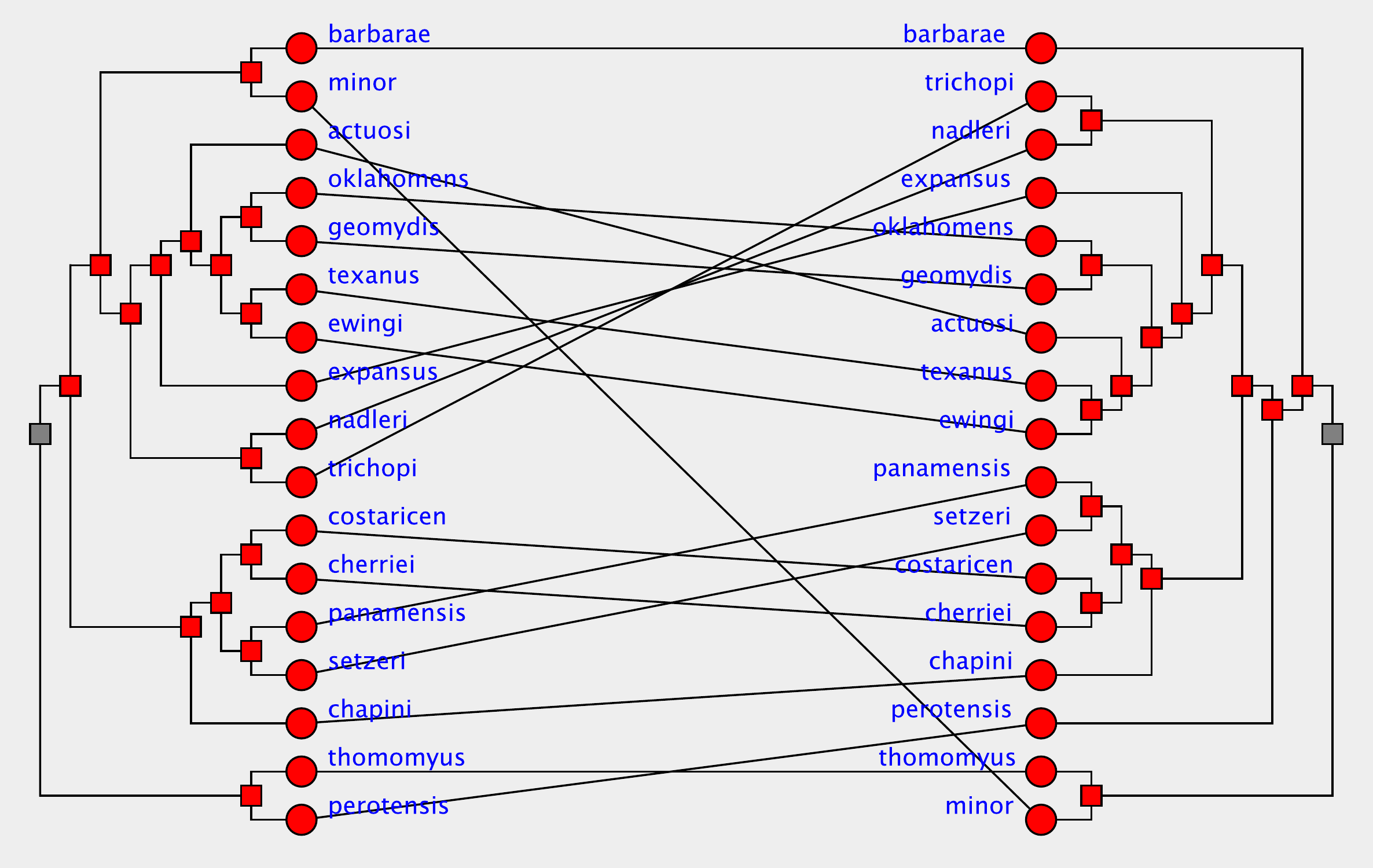}}
  \hfill
  \subfloat[optimal layout\label{sfg:untangled}]%
  {\includegraphics[width=.48\textwidth]{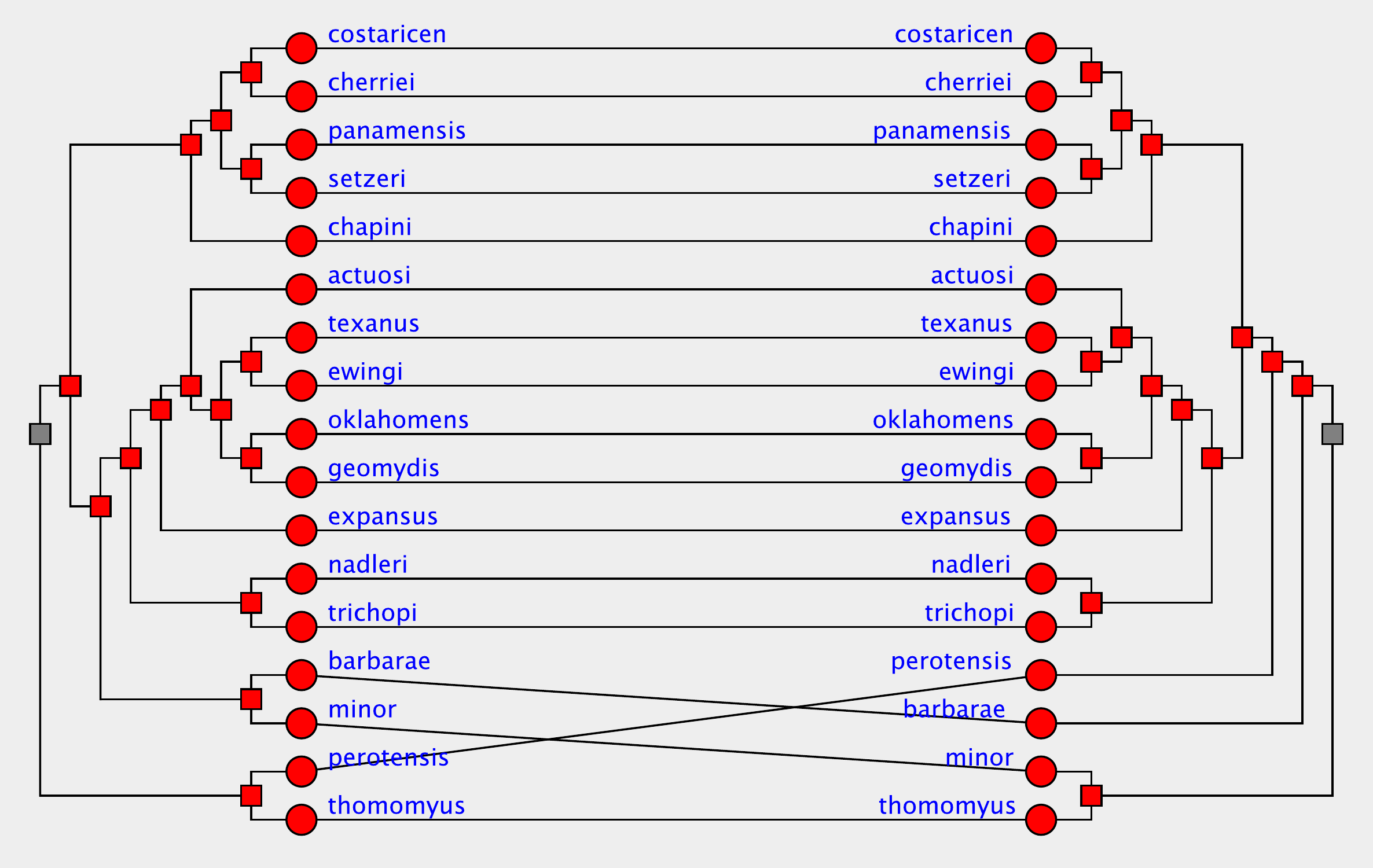}}
  \caption{A binary tanglegram of phylogenetic
    trees for lice of pocket gophers~\cite{hsvsdn-drmec-94}.}
  \label{fig:example}
  \vspace{-2ex}
\end{figure}

In this paper we are interested in evaluating the performance of two
recently suggested algorithms for drawing so-called
\emph{tanglegrams}~\cite{p-ttpcc-02}, that is, pairs of trees whose
leaf sets are in one-to-one correspondence.  The need to visually
compare pairs of trees arises in applications such as the analysis
of software projects, phylogenetics, or clustering.  In the first
application, trees may represent package-class-method hierarchies or
the decomposition of a project into layers, units, and modules.  The
aim is to analyze changes in hierarchy over time or to compare
human-made decompositions with automatically generated ones. Whereas
trees in software analysis can have nodes of arbitrary degree, trees
from our second application, that is, (rooted) phylogenetic trees,
are binary trees. This makes binary tanglegrams an interesting
special case, see \figurename~\ref{fig:example}. Hierarchical
clusterings, our third application, are usually visualized by a
binary tree-like structure called \emph{dendrogram}, where elements
are represented by the leaves and each internal node of the tree
represents the cluster containing the leaves in its subtree. Pairs
of dendrograms stemming from different clustering processes of the
same data can be compared visually using tanglegrams.

From the application
point of view it makes sense to insist that (a)~the trees under
consideration are drawn plane, that is, without edge crossings,
(b)~each leaf of one tree is connected by an \emph{inter-tree} edge to
the corresponding leaf in the other tree, and (c)~the number of
crossings among the inter-tree edges is minimized.  Following the
bioinformatics literature (e.g., \cite{p-ttpcc-02,lprvz-stapt-07}), we
call this the \emph{tanglegram layout} problem; Fernau et
al.~\cite{fkp-ctvcm-05} refer to it as \emph{two-tree crossing
  minimization}.\smallskip

\noindent\emph{Problem: (Tanglegram Layout (TL)).}
  Given a tanglegram \ttree{S,T} consisting of two rooted trees $S$
  and $T$ on $n$ leaves and a bijection between their leaf sets, find a
  tanglegram layout, that is, two plane drawings of $S$ and $T$, such that
  \begin{enumerate}
  \item the drawing of $S$ is to the left of the line $x=0$
    with all leaves on $x=0$;
  \item the drawing of $T$ is to the right of the line $x=1$
    with all leaves on $x=1$;
  \item the inter-tree edges are drawn as straight-line segments;
  \item the number of inter-tree edge crossings is minimum.
  \end{enumerate}
\smallskip

Given a tree~$T$, we say that a linear order of its leaves is
\emph{compatible} with~$T$ if for each node~$v$ of~$T$ the nodes in
the subtree of~$v$ form an interval in the order.  Note that TL is a
purely combinatorial problem.  In short, given two trees~$S$ and~$T$,
TL consists of finding an order~$\sigma$ of the leaves of~$S$
compatible with~$S$ and an order~$\tau$ of the leaves of~$T$
compatible with~$T$ such that the number of inversions between~$\tau$
and~$\sigma$ is minimum \cite{fkp-ctvcm-05,bbbnow-dbtha-08}.  Let the
\emph{crossing number} of a tanglegram \ttree{S,T} be the minimum
number of inter-tree edge crossings of any tanglegram layout of
\ttree{S,T}.

In the following we restrict our attention to \emph{binary}
tanglegrams such as, for example, pairs of phylogenetic trees or
clustering dendrograms. The restriction of TL to binary trees is
denoted as \emph{binary} TL.  After presenting related work
(Section~\ref{sec:related}), we introduce the algorithms that we want
to compare experimentally (Section~\ref{sec:algorithms}).  We first
sketch a recursive algorithm of Buchin et al.~\cite{bbbnow-dbtha-08}.
Then, in an algorithm engineering process we adapt their algorithm to
the needs of unbalanced trees in order to achieve better results.  We
apply branch-and-bound to speed up the improved variant.  Next, we
introduce hierarchy sort, a crossing-reduction heuristic used in the
visualization tool of Holten and van Wijk~\cite{hw-vchod-08}, and a
quadratic integer program that solves binary TL optimally.  Finally,
we provide a detailed description of the results of our experimental
comparison of these algorithms, see
Section~\ref{sec:experimental-results}.

\section{Related Work}\label{sec:related}

In graph drawing the so-called \emph{two-sided crossing minimization
  problem} (2SCM) is an important NP-hard problem that occurs when
computing layered graph layouts.  Such layouts have been introduced by
Sugiyama et al.~\cite{stt-mvuhs-81} and are widely used for drawing
hierarchical graphs. In 2SCM, vertices of a bipartite graph are to be
placed on two parallel lines (called \emph{layers}) such that for each
vertex on one line all its adjacent vertices lie on the other line.
As in TL the objective is to minimize the number of edge crossings
provided that edges are drawn as straight-line segments.  In one-sided
crossing minimization (1SCM) the order of the vertices on one of the
layers is fixed.  Even 1SCM is NP-hard~\cite{ew-ecdbg-94}.
J\"unger and
Mutzel~\cite{jm-tlscm-97} performed an experimental comparison of
exact and heuristic algorithms for both 1SCM and 2SCM. The main
findings were that for 1SCM the exact solution can be computed quickly
for up to 60 vertices in the free layer, and for 2SCM an iterated
barycenter heuristic is the method of choice for instances with more
than 15 vertices in each layer.

The main difference between TL and 2SCM is that in TL, the possible orders
of the leaves are limited to those that are compatible with the two
input trees. Furthermore the inter-tree edges are usually restricted
to be a matching of the leaves.  Dwyer and
Schreiber~\cite{ds-oloth-04} studied drawing a series of tanglegrams
in 2.5 dimensions, that is, the trees are drawn on a set of stacked
two-dimensional planes. They considered a one-sided version of
binary TL by fixing the layout of the first tree in the stack, and
then, layer-by-layer, computing an optimal compatible leaf order of
the next tree in $O(n^2 \log n)$ time each.  
Fernau et al.~\cite{fkp-ctvcm-05} showed
that binary TL is NP-hard and gave a fixed-parameter algorithm that runs in
$O^\star(c^k)$ time, where the $O^\star$-notation ignores polynomial
factors, $c$ is a constant that Fernau et al.\ estimate to
be $1024$, and $k$ is the minimum number of crossings in any drawing
of the given tanglegram.  They further showed that the problem can
be solved in $O(n \log^2 n)$ time if the leaf order of one tree is
fixed.  This improves the result of Dwyer and
Schreiber~\cite{ds-oloth-04}.  They also made the simple observation
that the edges of the tanglegram can be directed from one root to
the other.  Thus the existence of a planar drawing can be verified
using a linear-time upward-planarity test for single-source directed
acyclic graphs~\cite{bdmt-dupts-98}.  Later, apparently not knowing
these previous results, Lozano et al.~\cite{lprvz-stapt-07} gave a
quadratic-time algorithm for the same special case, to which they
refer as \emph{planar tanglegram layout}.  Recently, Buchin et
al.~\cite{bbbnow-dbtha-08} showed that binary TL remains NP-hard even if
both trees are complete binary trees. For this case they gave an
$O(n^3)$-time factor-2 approximation algorithm and a simple
$O^\star(4^k)$-time fixed-parameter algorithm, where $k$ is the
minimum number of crossings as before. Their approximation algorithm
is based on recursive splitting of the instance and can also be used
as a heuristic for general binary trees.  Holten and van
Wijk~\cite{hw-vchod-08} present a tanglegram visualization
tool for the comparison of two (not necessarily binary) trees that
uses local optimization to reduce inter-tree crossings
an edge-bundling technique to reduce visual clutter.

\section{Algorithms}\label{sec:algorithms}

In this section we describe the recursive splitting algorithm of
Buchin et al.~\cite{bbbnow-dbtha-08} and our improved variant of it,
then the algorithm hierarchy sort of Holten and van
Wijk~\cite{hw-vchod-08}, and finally a simple integer quadratic
program (IQP) that provides us with exact solutions for the
experimental comparison that follows in
Section~\ref{sec:experimental-results}.

\subsection{Recursive Splitting Algorithm}\label{sec:recurs-splitt-algo}

\begin{wrapfigure}[11]{r}{5cm}
  \centering
  \vspace*{-4.5ex}
  \includegraphics{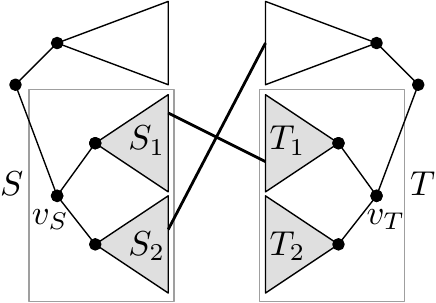}
  \caption{A subinstance \ttree{S,T} with a current-level crossing.}
  \label{fig:subinstance}
\end{wrapfigure}

The main idea behind the recursive splitting algorithm is to
recursively consider for an instance \ttree{S,T} the four possible
orders of the two subtrees $S_1,S_2$ of $S$ and
$T_1,T_2$ of $T$ below the roots $v_S$ and $v_T$ of $S$ and $T$ as in
\figurename~\ref{fig:subinstance}.  Each order gives rise to a certain
number of crossings at that level of the recursion (called
\emph{current-level} crossings), which is added to the number of
crossings of both recursively solved subproblems induced by that order
(called \emph{lower-level} crossings). 
Each current-level 
crossing has
the property that it can be removed by swapping the subtrees of $v_S$
or $v_T$. For example the crossing depicted in
\figurename~\ref{fig:subinstance} can be removed by swapping the
subtrees of $v_S$ and placing $S_2$ above $S_1$. Of course such a swap
generally introduces other current-level crossings. The minimum of the
four possibilities will be returned to the previous level of the
recursion. The two subproblems that arise from each recursive split
are not independent.  Nevertheless, they are treated independently by
the algorithm. This obviously introduces an error with respect to the actual
number of crossings, which, for the case of \emph{complete} binary
trees, can be bounded by the number of crossings in an optimal
solution~\cite{bbbnow-dbtha-08}.  For complete binary trees the
recursive algorithm thus yields a 2-approximation.  Obviously, the
depth of the recursion equals the minimum height~$h$ of the two
trees. The recursion tree is of size $O(8^h)$ since each instance
starts eight recursive calls (two for each of the four subtree
arrangements). %
The computation of all current-level crossings is done in $O(4^h n)$
time, resulting in a total running time of $O(8^h + 4^h n)$.
For
complete trees with $h = \log n$ this resolves to $O(n^3)$ time.

\begin{figure}[tb]
  \centering
  \subfloat[optimal layout: 1 crossing]%
  {\includegraphics[width=.36\textwidth]{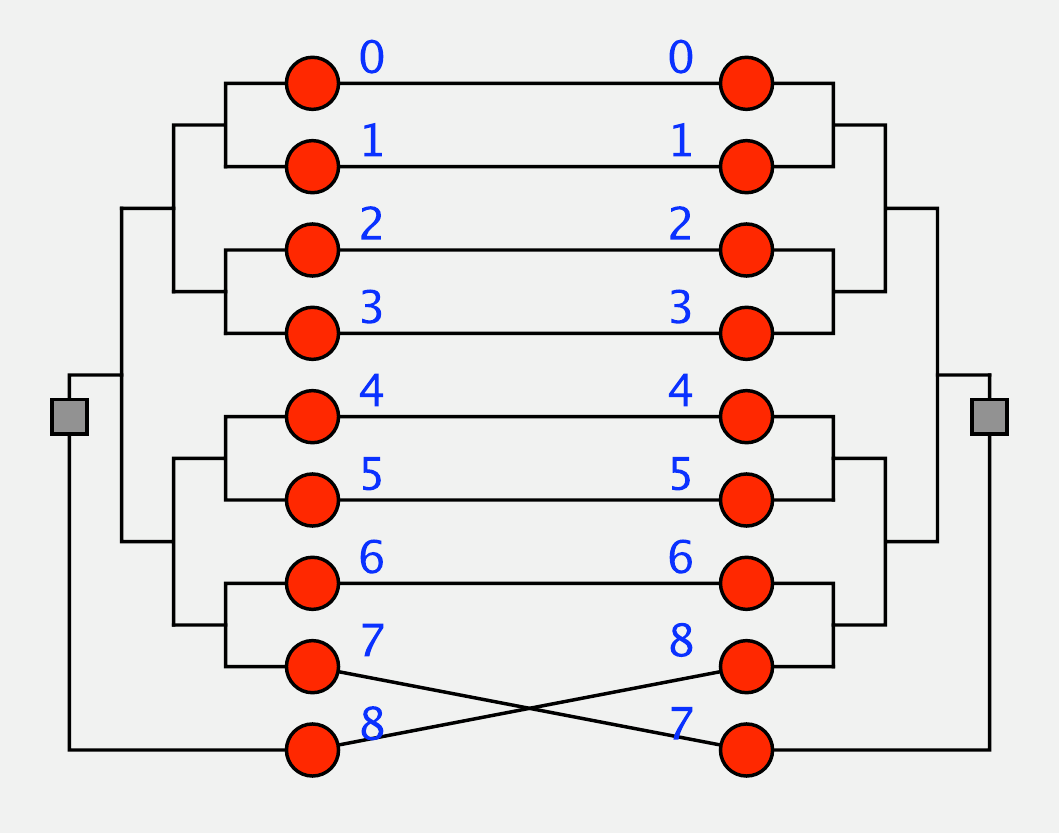}}
  \hfil
  \subfloat[heuristic layout: 14 crossings\label{fig:alg-bad-14}]%
  {\includegraphics[width=.36\textwidth]{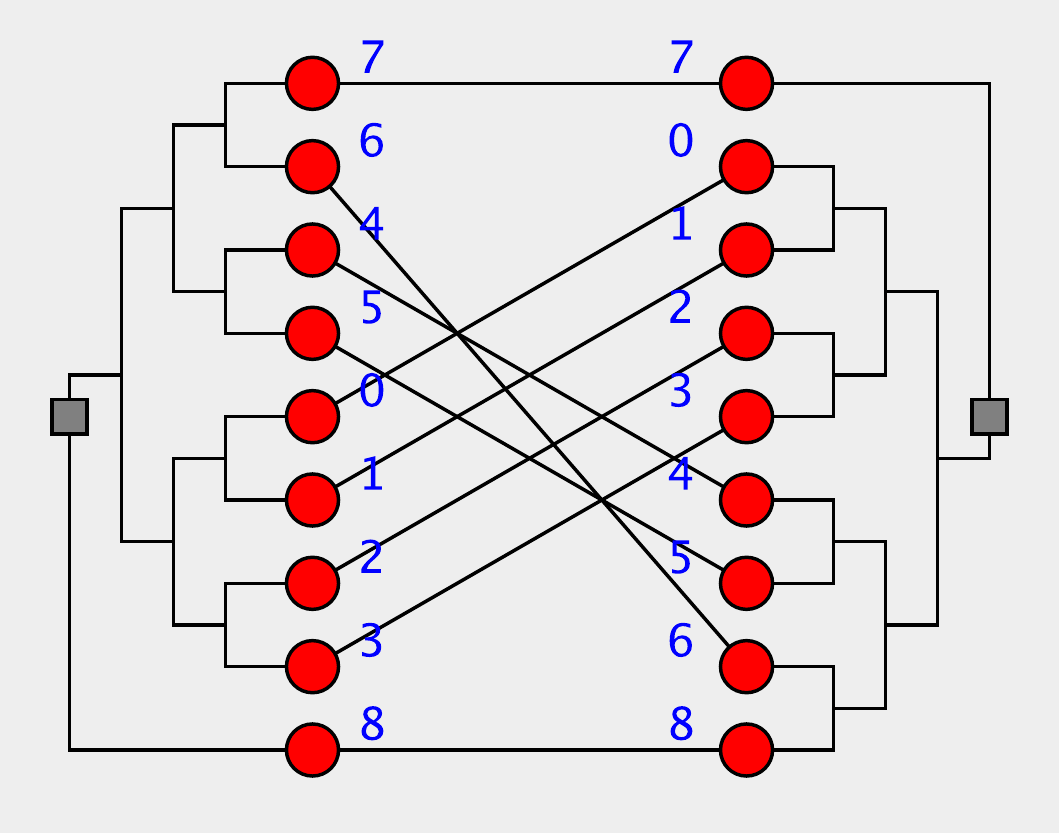}}
  \caption{Example of a binary tree for which the original heuristic performs badly.}
  \label{fig:alg-bad}
\end{figure}

In applications most binary TL instances do not consist of \emph{complete} binary
trees. The above recursive algorithm can be applied to any pair of
binary trees as a heuristic but an approximation guarantee cannot be
given any more. Under the Unique Games Conjecture a constant-factor
approximation does not even exist for general binary
trees~\cite{bbbnow-dbtha-08}. The original algorithm always divides an
instance into an upper and a lower subinstance, that is, the two problems
\ttree{S_1,T_1} and \ttree{S_2,T_2} in the example of
\figurename~\ref{fig:subinstance}. For unbalanced trees this can lead
to an unnecessarily high number of ignored crossings as
\figurename~\ref{fig:alg-bad} shows.  The original algorithm aligns
the leaves (nodes~7 and~8) attached directly to the roots %
since this causes no current-level
crossings. All 14 crossings in \figurename~\ref{fig:alg-bad-14} are
crossings that the algorithm does not take into account. A small
modification of our algorithm weakens this effect (and yields the
optimum solution in the given example). Instead of always dividing
into an upper and a lower subinstance, we can also consider dividing
into the two \emph{diagonal} subinstances \ttree{S_1,T_2} and
\ttree{S_2,T_1} in the example of
\figurename~\ref{fig:subinstance}. The improved algorithm always
selects among the two possible splits the one that has the higher
total number of edges between its two subinstances. This modification
not only improves the algorithm performance, but it also allows us to
precompute in $O(n^2 h)$ time all required numbers of current-level
crossings for a constant-time lookup.
Thus the total running time reduces to $O(8^h + n^2 h)$, which still
equals $O(n^3)$ for complete trees.  Interestingly, it can be proved
(omitted here) that the approximation factor of~2 still holds for this
modified algorithm in the case of complete trees.

In our most refined implementation of the improved algorithm we
additionally make use of a branch-and-bound technique in order to
prune large parts of the search tree as early as possible.  This
considerably sped up the naive implementation, see
Section~\ref{sec:runn-time-perf}. Let's consider an instance of the
problem with roots~$v_{S'}$ and~$v_{T'}$. Instead of computing the
number of crossings for all four possible arrangements of the
respective subtrees of~$v_{S'}$ and~$v_{T'}$ we first consider the one
that yields the lowest number of current-level crossings and
recurse. This gives us an initial upper bound on the number of
crossings once the leaf level is reached. Now at each level we can
immediately prune the respective parts of the search tree for those
arrangements that exceed this upper bound. The rest of the search tree
is examined further, and each time a better solution is found the
upper bound is updated accordingly.

\subsection{Hierarchy Sort}\label{sec:hier-sort-algo}

\begin{figure}[tb]
  {\includegraphics[width=\textwidth]{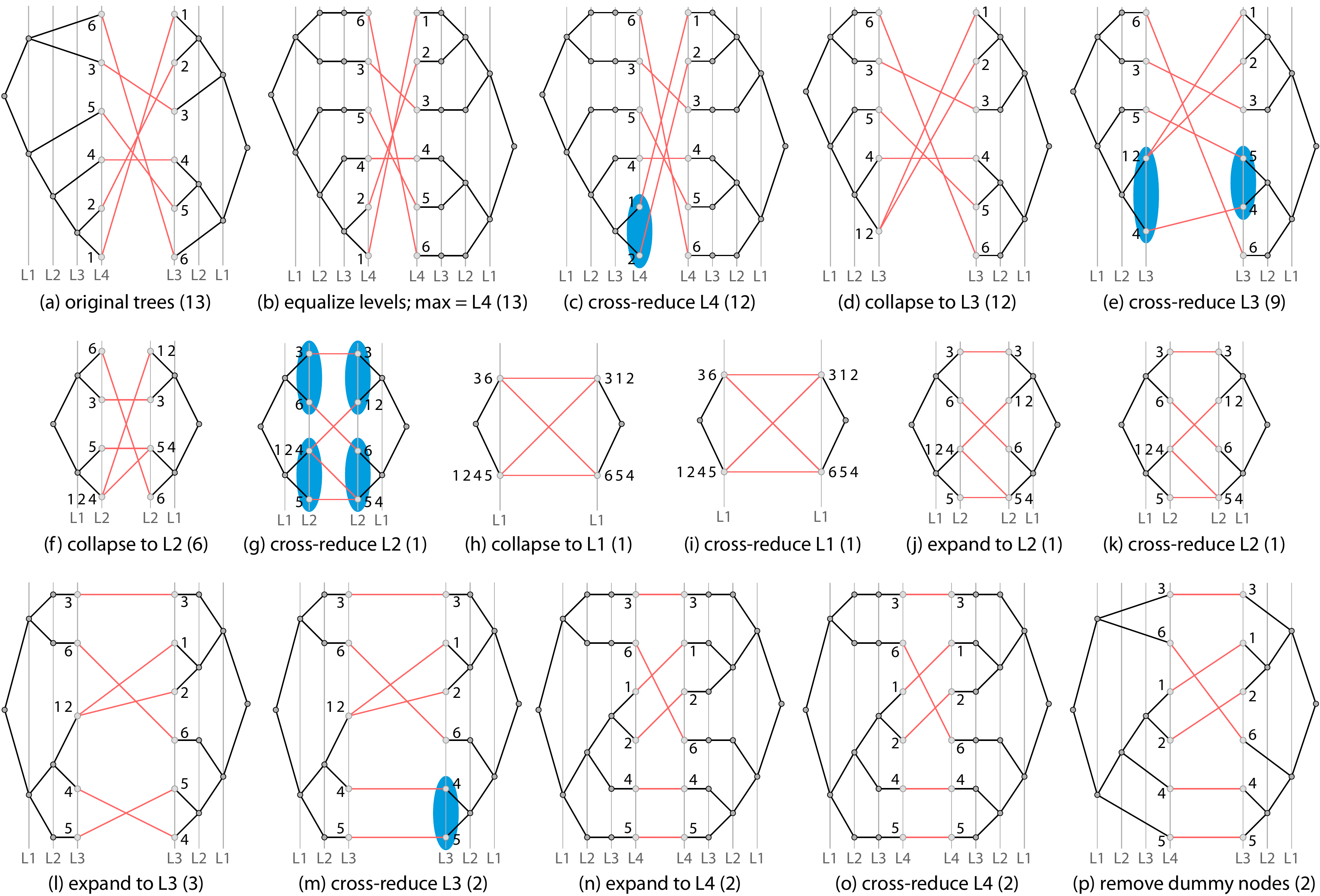}}
  \caption{Step-by-step crossing reduction (CR) %
    using hierarchy sort. Nodes that are swapped during CR
    are encircled. The numbers of %
    crossings after each step are given in parentheses.}
  \label{fig:hierarchysorting}
\end{figure}

The algorithm hierarchy sort of Holten and van Wijk~\cite{hw-vchod-08}
performs a number of collapse-and-expand cycles on both trees of the
binary tanglegram.  During each step of these cycles, the well-known
barycentric method of Sugiyama et al.~\cite{stt-mvuhs-81} for 1SCM is
used by successively fixing one tree, optimizing
the leaf order of the other, and then changing the trees' roles until no
further crossing reduction is possible. 
We illustrate the algorithm using the example in
Fig.~\ref{fig:hierarchysorting}. 

Figure~\ref{fig:hierarchysorting}a shows a binary tanglegram with 13
inter-tree edge crossings.  Figures~\ref{fig:hierarchysorting}b
to~\ref{fig:hierarchysorting}p illustrate the hierarchy sorting
algorithm using one full collapse-and-expand cycle. Since crossings
are to be reduced on \emph{corresponding} levels in the two trees,
the numbers of levels of the two trees need to equalized.  This is
done by introducing dummy nodes that bring all leaves to the lowest
level.  In our example, this results in a
tanglegram consisting of two four-level binary trees, see
\figurename~\ref{fig:hierarchysorting}b.

Crossing reduction (CR) is now performed per level by employing the
barycentric method on corresponding levels. Due to the hierarchical
structure of the data that we consider, only nodes having the
same parent may be swapped.  Nodes that
are swapped during CR are encircled in
\figurename~\ref{fig:hierarchysorting}.

After having completed CR at the lowest level, we move up one
level in both trees.  We do this by \emph{collapsing} both levels,
that is, by contracting all edges ending in leaves (see the step
from \figurename~\ref{fig:hierarchysorting}c
to~\ref{fig:hierarchysorting}d, for example).  CR can now proceed.
Collapsing and CR are repeated until the levels below the roots are
reached (\figurename~\ref{fig:hierarchysorting}i).  At this point,
the process is reversed and levels are expanded again (with
in-between CR) until the leaf levels are reached.  This is
illustrated in Figs.~\ref{fig:hierarchysorting}j
to~\ref{fig:hierarchysorting}o. Such collapse-and-expand cycles are 
repeated until the number of crossings does not decrease any
further.

A last step remains: the original number of levels in both
hierarchies needs to be restored, that is, all dummy
nodes are contracted (see \figurename~\ref{fig:hierarchysorting}p).
In our example the hierarchy sorting algorithm has reduced the
number of crossings from~13 to~2.

The asymptotic running time of this algorithm depends of course on the
number~$N$ of collapse-and-expand cycles and the maximum number~$N'$
of executions of the linear-time barycentric heuristic on each level.
In our experiments (see Section~\ref{sec:experimental-results}) it
turned out that in all instances we had $N \le 2$ and $N' \le 2$.
Under the condition that both $N$ and $N'$ are constants, hierarchy
sort runs in $O(n \cdot H)$ time, where $H$ is the maximum height of
the two trees. In the case of complete trees $H= \log n$, and the
running time is $O(n \log n)$.

\subsection{Integer Quadratic Program}\label{sec:int-quad-prog}

For the IQP we introduce a binary variable~$x_u$ for each inner node
of $S \cup T$.  If~$x_u=1$, the two subtrees of~$u$ change their
order with respect to the input drawing, otherwise the order of the
input drawing is kept.  Let~$ab$ and~$cd$ be two inter-tree edges
with~$a,c \in S$ and $b,d \in T$.  Let $v \in S$ and $w \in T$ be
the lowest common ancestors of the leaves~$a$ and~$c$, and of~$b$
and~$d$, respectively.  Assume that~$ab$ and~$cd$ cross each other
in the original drawing.  Then~$ab$ and~$cd$ cross each other in the
solution encoded by the IQP if and only if $x_u \cdot x_v = 1$ or
$(1-x_u) \cdot (1-x_v) = 1$.  Otherwise, if~$ab$ and~$cd$ do not
cross each other originally, they will cross in the solution encoded
by the IQP if and only if $x_u \cdot (1-x_v) = 1$ or $(1-x_u) \cdot
x_v = 1$. Thus the total number of edge crossings can be expressed
as the sum of these products for all pairs of edges.  The IQP
minimizes this sum as its objective function. No further
constraints, apart from the variables being binary, are necessary.

\section{Experimental Results}\label{sec:experimental-results}

The recursive splitting algorithms were written in Java~1.5 and
executed in SuSE Linux~9.3 running on an AMD Opteron~248 2.2 GHz
system with 4 GB RAM.  The hierarchy sorting algorithm was implemented
in Delphi~7.0 and executed in Windows XP on an Intel Pentium~4 2.8 GHz
system with 1 GB RAM.  The quadratic program was solved with the
mathematical programming software CPLEX~9.1 running on the above Linux
system.

\subsection{Data}

We generated four sets (A--D) of random tanglegrams. Set~A contains
ten pairs of complete binary trees with random leaf orders for each
$n = 16, 32, \ldots, 256$.  In set~B we simulated tree mutations by
starting with two identical complete binary trees and then randomly
swapping the positions of up to 20\% of the leaves of one tree. This
is done as follows: we first pick a leaf uniformly at random and
then iteratively climb up the tree with probability 0.75 in each
step. From the node thus reached we climb back down and flip a coin
at each node to choose its left or right child until we reach
another leaf. This leaf and the leaf picked in the beginning are
swapped.  Thus the probability of two leaves being swapped decreases
with their distance in the tree. Set~C contains ten pairs of general
binary trees for each $n = 20, 40, \ldots, 200$. The trees are
constructed from a set of nodes, initially containing the $n$
leaves, by iteratively joining two random nodes in a new parent node
that replaces its children in the set. This process generates trees
that resemble phylogenetic trees or clustering dendrograms. Set~D is
similar to set~C but again in each tanglegram the second tree is a
mutation of the first tree, where up to 10\% of the leaves can swap
positions as done in set~B and up to 25\% of the subtrees can
reattach to another edge.  This edge is selected in a random walk
starting at the subtree's old position. The walk continues with
probability 0.75 and chooses the left or right edge by tossing a
coin.  Trees in this set are of interest since real-world
tanglegrams often consist of two related and rather similar trees. The
average crossing numbers of the trees in sets~A--D are given in
\figurename~\ref{fig:crnumbers-random} in the appendix. 

Our real-world examples comprise three sets (E--G) of
tanglegrams. Set~E contains six pairs of dendrograms of a
hierarchically clustered social network based on email communication
of 21 subjects~\cite{ggw-lvavl-08}. Sets~F and~G contain six and ten
pairs of phylogenetic trees for 15 species of pocket gophers and 17
species of lice, respectively~\cite{hsvsdn-drmec-94}. 
(\figurename~\ref{fig:example} shows a tanglegram in set~G.)
While the email tanglegrams have between 23 and 45 crossings
in an optimal solution, the phylogenetic trees can be drawn with at
most two crossings, most of them even without any crossings.

\subsection{Performance}

In the following we denote the original recursive 
algorithm of Buchin et al.~\cite{bbbnow-dbtha-08} by \recsplit\ and
our modification for unbalanced trees by 
\recimproved. The algorithm of Holten and van Wijk is 
\hiersort.  Let $n$ be the size of an instance, that is, the number of
leaves per tree. %

To each tanglegram we applied the three algorithms and the IQP, and
recorded the crossing numbers $c_i$ in their respective solutions for
$i =$ \recsplit, \recimproved, \hiersort.  We then computed for each
tanglegram the performance ratios $(c_i + 1)/(c_\mathrm{opt} + 1)$,
where $c_\mathrm{opt}$ denotes the optimal number of crossings
obtained from solving the IQP. Note that we add one to both crossing
numbers in order to have a well-defined ratio also for crossing-free
instances.

\begin{figure}[tbp]
  \centering
  \includegraphics[width=.49\textwidth]{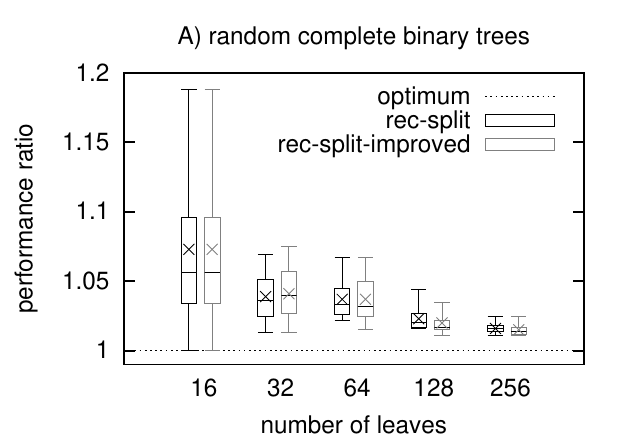}
  \hfill
  \includegraphics[width=.49\textwidth]{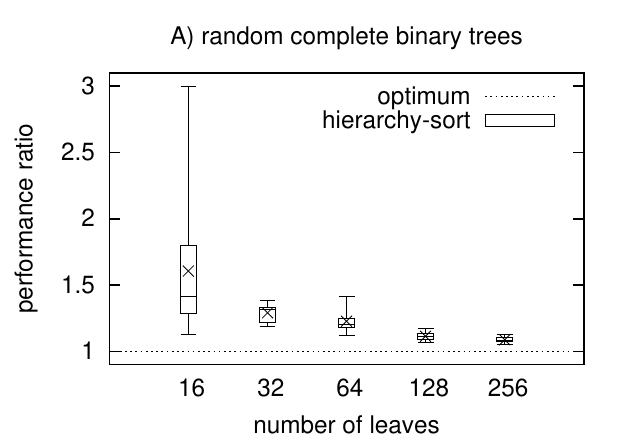}\\
  \includegraphics[width=.49\textwidth]{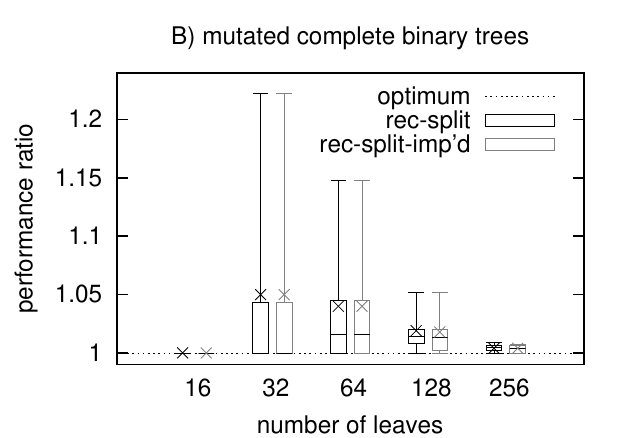}
  \hfill
  \includegraphics[width=.49\textwidth]{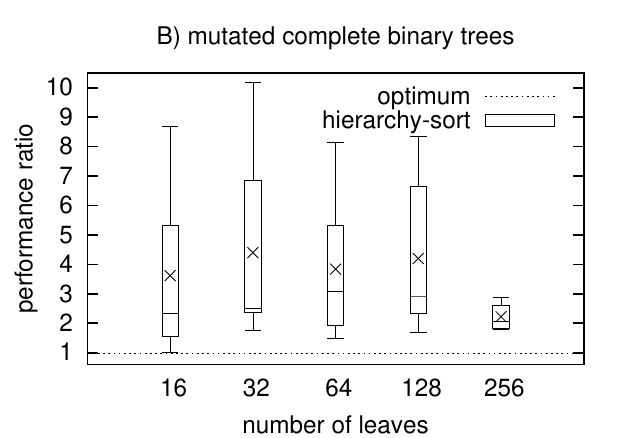}\\
  \includegraphics[width=\textwidth]{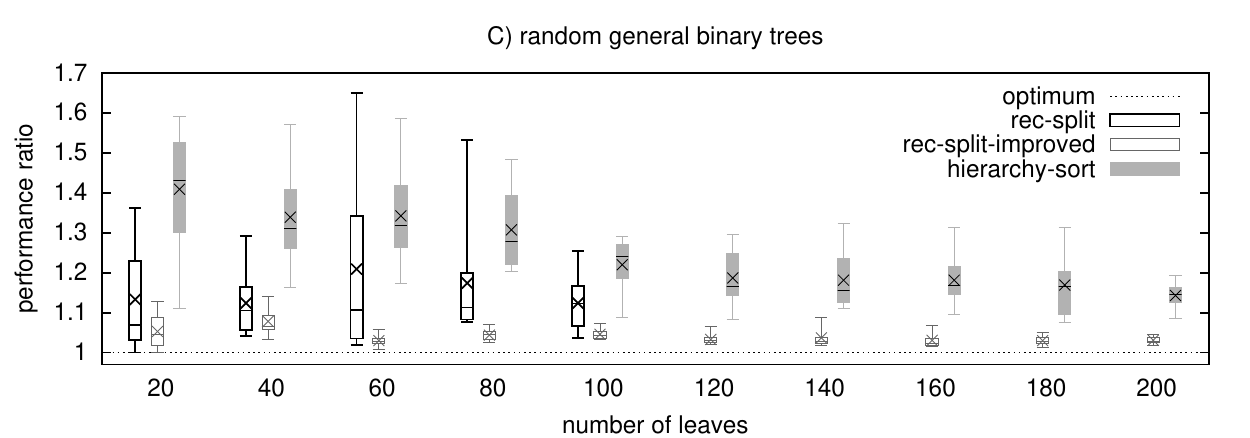}\\
  \includegraphics[width=\textwidth]{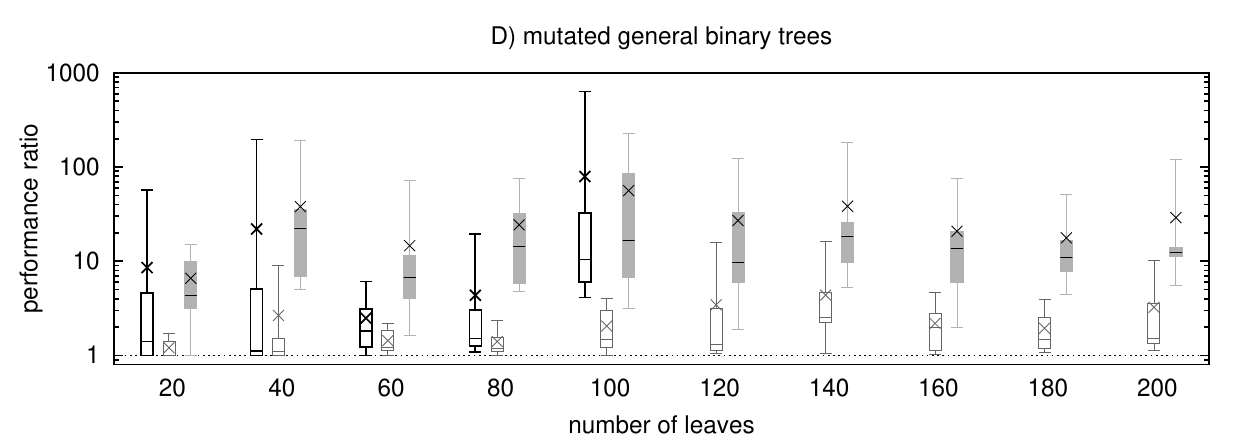}
  \caption{Performance ratios of the three algorithms \recsplit,
    \recimproved, and \hiersort.  The boxplots show medians, first and
    third quartiles, minimum and maximum values. Arithmetic means are
    indicated by crosses.}
  \label{fig:ratio}
\end{figure}

The results for sets~A--D are shown in \figurename~\ref{fig:ratio}.
For complete binary trees (sets~A and~B) \recsplit\ and \recimproved\
achieved similar performance ratios that tend to~1 as the size of the
trees grows.  Recall that on these complete instances both algorithms
are 2-approximations. On average both methods performed slightly
better on mutated trees (B) than on random trees (A).  In several
cases the instances in set~B could be solved optimally. The average
performance ratio of \hiersort\ was slightly worse for the random trees
of set~A and drastically worse for the mutated trees of set~B with
average values between 2.23 and 4.4 in comparison to values between 1
and 1.05 for the recursive algorithms.  Furthermore, \hiersort\
performed better on random trees rather than on mutated trees. Note
that the absolute number of crossings is lower for mutated trees; thus
a difference of only 1 or 2 to the optimum can already lead to
relatively large ratios for small $n$.

For general binary trees the performance ratios of \recsplit\ and
\recimproved\ are no longer upper-bounded by 2 but at least for random
trees (C) the ratios were on average well below~2. As expected,
\recimproved\ outperformed \recsplit\ due to the modification for
unbalanced trees. Algorithm \recsplit\ attained performance ratios
close to 1 for most random instances but it had some outliers as
well. The solutions of \recimproved\ were not only closer to the
optimum, they also spread much less. Note that due to excessive
computation times of several hours we did not record the results of
\recsplit\ for $n \ge 110$. The hierarchy sorting algorithm yielded
results that were clearly inferior to those of the recursive
algorithms. But it still achieved performance ratios below~1.2 as~$n$
grows. In general, the behavior of both \recimproved\ and \hiersort\
did not differ much between sets~A and~C and thus the completeness of
the trees seems of low impact on the solution quality. In contrast
\recsplit\ performed worse on general trees than on complete trees.

For the mutated trees of set~D with relatively fewer crossings in the
optimal solution we use a logarithmic scale for the performance
ratio. The results were generally worse and spread a lot more for all
three algorithms, but still \recimproved\ had the best performance
with the upper quartile of the ratios mostly below~3. 
On the other
hand the upper quartile of \hiersort\ reached 
values mostly above~20 and up
to~85 for $n=100$. Algorithm \recsplit\ reached performance ratios
between those of the other two algorithms. For $n \ge 90$ we stopped
considering \recsplit\ since its computation times became again too
high.

\begin{wrapfigure}[12]{r}{5cm}
  \centering
  \vspace*{-5ex}
  \includegraphics[scale=.9,viewport=17 4 169 119,clip]{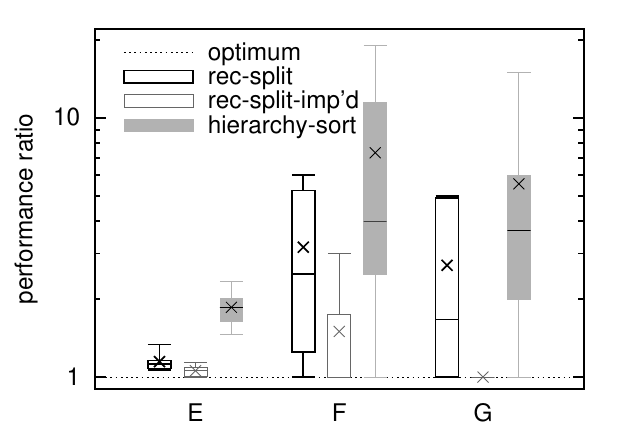}
  \vspace*{-1ex}
  \caption{Performance ratios for real-world examples.}
  \label{fig:real_world}
\end{wrapfigure}
The relative performance of the algorithms for random
non-complete trees was confirmed by the results for the sets~E--G of
real-world examples, see Fig.~\ref{fig:real_world}.
For the clustering data of set~E with an average
of 33.5 crossings \recimproved\ reached an average performance ratio
of 1.06, \recsplit\ was slightly
worse with 1.15, and \hiersort\ had an
average ratio of 1.86. The phylogenetic data of sets~F and~G can often
be drawn without crossings and thus have average crossing numbers of
only 0.17 and 0.7, respectively. This explains the relatively high
performance ratios. Still, \recimproved\ found an optimum layout for
four of the six examples in set~F and solved all ten instances in set~G
optimally.

\subsection{Running Time}\label{sec:runn-time-perf}

\begin{figure}[tb]
  \centering
  \includegraphics[width=.49\textwidth]{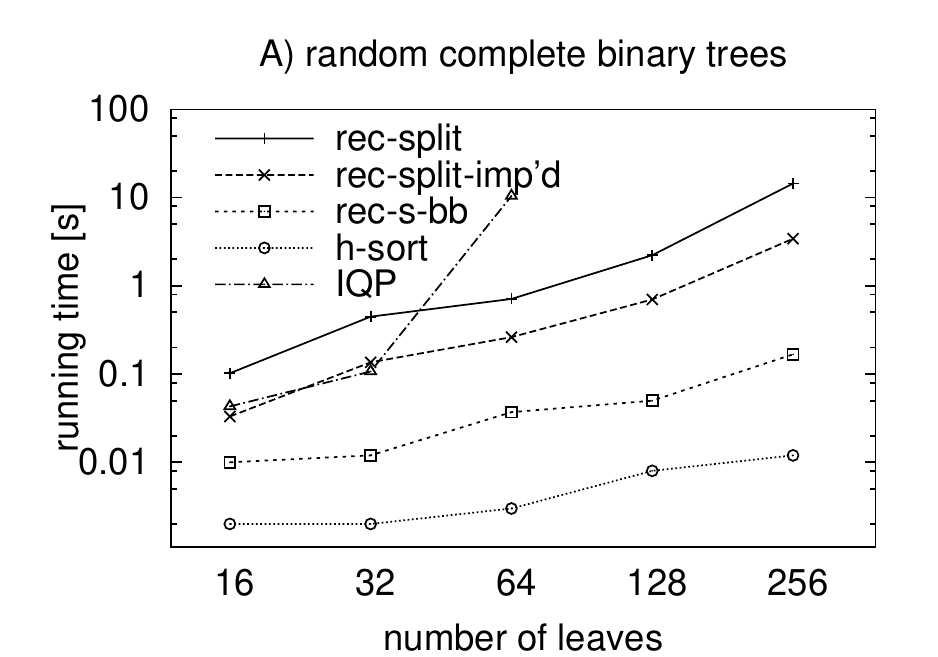}
  \hfill
  \includegraphics[width=.49\textwidth]{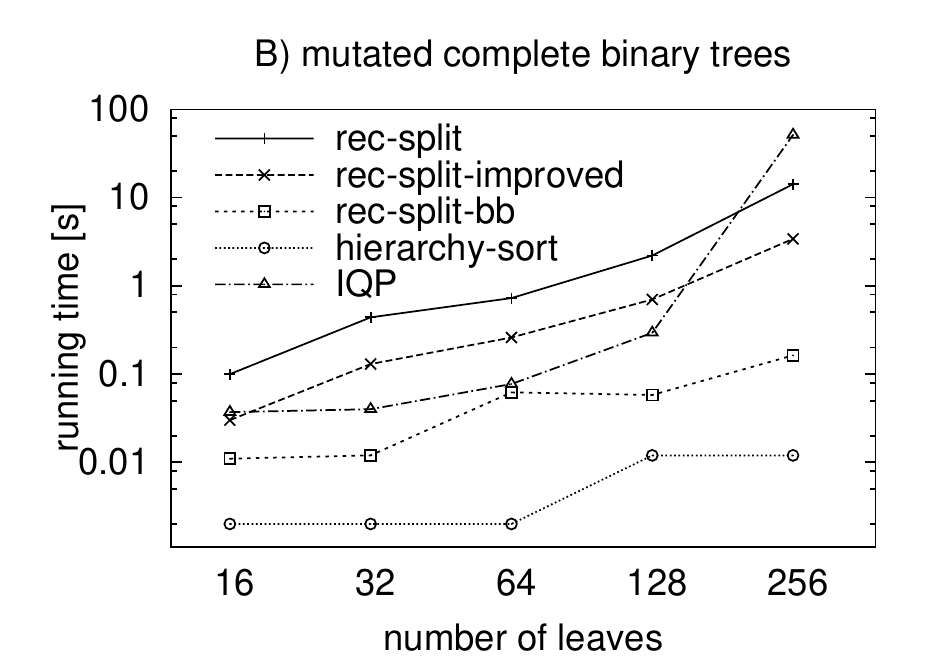}\\
  \includegraphics[width=.49\textwidth]{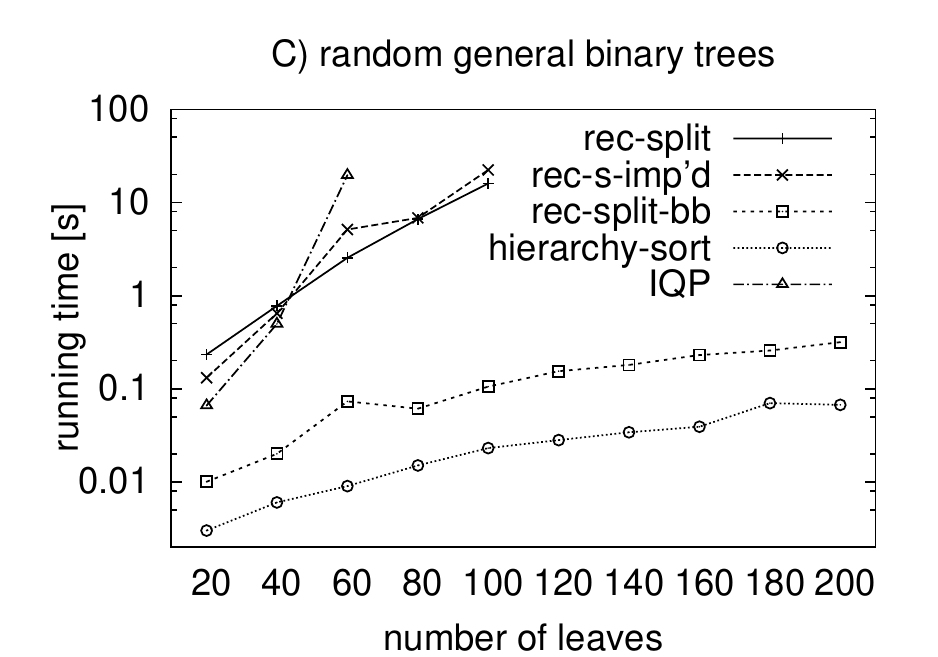}
  \hfill
  \includegraphics[width=.49\textwidth]{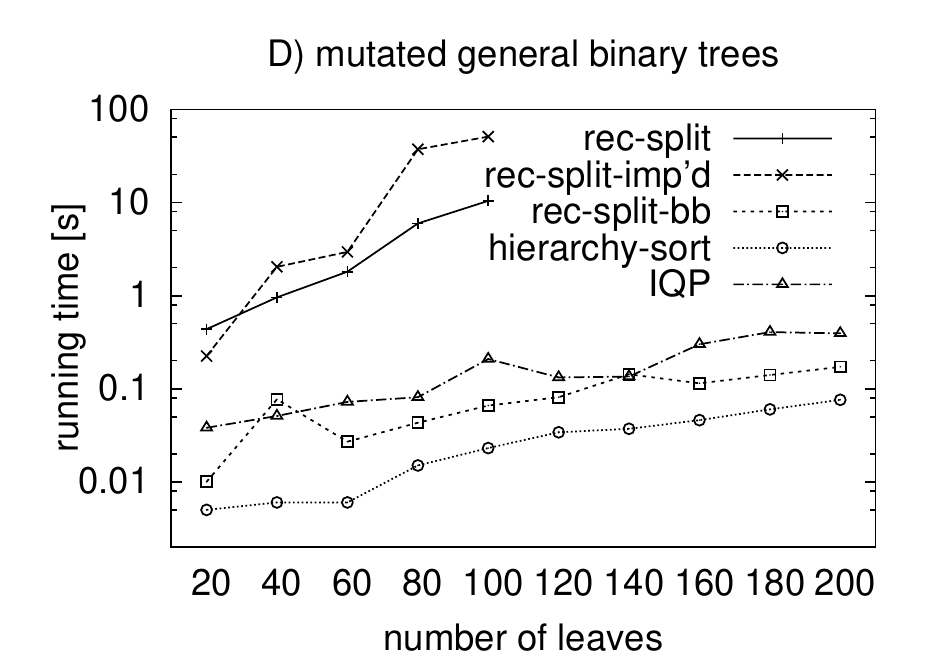}
  \caption{Median running times of the algorithms \recsplit, \recimproved,
    \recbb, \hiersort, and IQP (in seconds).  Note the use of log scales.}
  \label{fig:times}
  \vspace{-4ex}
\end{figure}

Although the number of crossings is the main aspect to assess the
quality of TL algorithms, their running time is also
important---especially if the layouts are to be produced
interactively. Figure~\ref{fig:times} shows plots of the running times
of \recsplit, \recimproved, \recbb\ (the branch-and-bound
implementation of \recimproved), \hiersort, and the IQP for all four
classes of random tanglegrams.  Note the use of log scales.  
Recall that \hiersort\ was written in
Delphi instead of Java and executed on a different system. Hence the
absolute running times of \hiersort\ are to be taken with a grain of
salt. 

By far the fastest algorithm in all our examples was \hiersort,
which took at most 12 ms for complete trees and less than 90 ms for
arbitrary trees. No difference in terms of the running time could be
seen between random pairs and mutated pairs of trees. In contrast,
the measured running times of the branch-and-bound algorithm \recbb\
were about ten times higher for complete trees and between four and
six times higher for general trees. Still, the median running time
of \recbb\ was less than 360 ms for all instances. In the direct
comparison of random~(C) and mutated~(D) pairs of non-complete trees
with the same number of leaves, the random pairs, which have a much
larger crossing number, required between 50 and 100\% more computation
time.

The naive recursive implementations of \recsplit\ and \recimproved\
were far slower than \hiersort\ and \recbb. For complete binary
trees they both grow at a cubic rate in $n$, but \recimproved\ is
about three times faster than \recsplit. This is due to the additional
$O(n^2 h)$-time 
preprocessing step mentioned in Section~\ref{sec:recurs-splitt-algo}.
Both algorithms
are not influenced by the class of the complete trees (A or B) as to
be expected from their definitions. For general binary trees their
running times quickly grew up to several hours, at least for some of
the instances, which is due to the fact that the
running time is exponential in the tree height. Thus
complete trees with 256 leaves could be solved as fast as
some random trees with only 50 leaves. The better performance in
terms of crossings of \recimproved\ for unbalanced trees is paid for
by higher running times in comparison to \recsplit.

Interestingly, layouts for pairs of mutated and thus rather similar
trees took much more time to compute than layouts for two random
trees. One explanation is that for a subinstance \ttree{S',T'} the
smaller of the heights of $S'$ and $T'$ determines the recursion
depth. Thus for two similar trees with similar heights the recursion
depth will be larger on average than for two random trees with fairly
different heights. It is also noteworthy that the running time of
\recbb\ in our experiments was dominated by the above mentioned $O(n^2
h)$-time preprocessing step, which unlike the recursive algorithms
does not depend exponentially on the height.

Finally, we look at the running times of the IQP. In contrast to the
recursive algorithms the running time of the IQP is independent of the
height (and thus the completeness) of the trees. Rather it is the
value of the objective function, that is, the %
crossing number, %
that influences the solution time. Therefore all
mutated trees, which have relatively small crossing numbers, could be
solved optimally within the time limits of ten minutes. On the other
hand for random tanglegrams optimality could only be proven for $n \le
64$.
For larger instances a slowly increasing %
gap between the best found integer solution and the fractional
solution remained, see Fig.~\ref{fig:crnumbers-random}~(top).

\section{Conclusions}

The experimental evaluation shows that in terms of crossing
reduction our improvement of the recursive splitting algorithm of Buchin et
al.~\cite{bbbnow-dbtha-08} clearly has the best performance for all
instances that were included in the tests.  Moreover, our
branch-and-bound implementation is fast enough (less than 0.4
seconds for trees with 200 leaves each) to be used interactively.
Thus it is the method of choice for drawing binary tanglegrams with
up to a few hundred leaves. Still, in terms of running time the
hierarchy sorting heuristic of Holten and van
Wijk~\cite{hw-vchod-08} outperforms the recursive splitting
algorithm; it can thus also be used for very large trees if the number
of crossings is not the main optimization criterion.  Also, it is
currently the only method that can draw non-binary tanglegrams.  For 
medium-sized tanglegrams that consist of two similar trees and thus
have a rather small crossing number it is worth to give it a try and
solve the very simple integer quadratic program
to obtain the optimal solution---often this takes but a few seconds.%

%
%
%


\begin{thebibliography}{10}

\bibitem{bdmt-dupts-98}
P.~Bertolazzi, G.~{Di Battista}, C.~Mannino, and R.~Tamassia.
\newblock Optimal upward planarity testing of single-source digraphs.
\newblock {\em SIAM J. Comput.}, 27(1):132--169, 1998.

\bibitem{bbbnow-dbtha-08}
K.~Buchin, M.~Buchin, J.~Byrka, M.~N\"ollenburg, Y.~Okamoto,
R.~I.~Silveira, and A.~Wolff.
\newblock Drawing binary tanglegrams: Hardness, approximation, fixed-parameter
  tractability.
\newblock Available at http://arxiv.org/abs/0806.0920

\bibitem{ds-oloth-04}
T.~Dwyer and F.~Schreiber.
\newblock Optimal leaf ordering for two and a half dimensional phylogenetic
  tree visualization.
\newblock {\em Proc. Australasian
  Sympos. Inform. Visual. (InVis.au'04)}, volume~35 of {\em CRPIT}, pages
  109--115. Australian Computer Society, 2004.

\bibitem{ew-ecdbg-94}
P.~Eades and N.~Wormald.
\newblock Edge crossings in drawings of bipartite graphs.
\newblock {\em Algorithmica}, 10:379--403, 1994.

\bibitem{fkp-ctvcm-05}
H.~Fernau, M.~Kaufmann, and M.~Poths.
\newblock Comparing trees via crossing minimization.
\newblock In {\em Proc. 25th Intern. Conf.
  Found. Softw. Techn. Theoret. Comput. Sci. (FSTTCS'05)}, volume 3821 of {\em
  Lecture Notes Comput. Sci.}, pages 457--469, 2005.

\bibitem{ggw-lvavl-08}
R.~G{\"o}rke, M.~Gaertler, and D.~Wagner.
\newblock {LunarVis -- Analytic Visualizations of Large Graphs}.
\newblock In {\em Proc. 15th
  Internat. Sympos. Graph Drawing (GD'07)}, volume 4875 of {\em Lecture Notes
  Comput. Sci.}, pages 352--364. Springer-Verlag, 2008.

\bibitem{hsvsdn-drmec-94}
M.~S. Hafner, P.~D. Sudman, F.~X. Villablanca, T.~A. Spradling, J.~W. Demastes,
  and S.~A. Nadler.
\newblock Disparate rates of molecular evolution in cospeciating hosts and
  parasites.
\newblock {\em Science}, 265:1087--1090, 1994.

\bibitem{hw-vchod-08}
D.~Holten and J.~J.~van Wijk.
\newblock Visual comparison of hierarchically organized data.
\newblock In {\em Proc. 10th Eurographics/IEEE-VGTC Sympos. Visualization
  (EuroVis'08)}, pages 759--766, 2008.

\bibitem{jm-tlscm-97}
M.~J\"unger and P.~Mutzel.
\newblock 2-layer straightline crossing minimization: Performance of exact and
  heuristic algorithms.
\newblock {\em J. Graph Algorithms Appl.}, 1(1):1--25, 1997.

\bibitem{lprvz-stapt-07}
A.~Lozano, R.~Y. Pinter, O.~Rokhlenko, G.~Valiente, and M.~Ziv-Ukelson.
\newblock Seeded tree alignment and planar tanglegram layout.
\newblock In {\em Proc. 7th Internat.
  Workshop Algorithms Bioinformatics (WABI'07)}, volume 4645 of {\em Lecture
  Notes Comput. Sci.}, pages 98--110. Springer-Verlag, 2007.

\bibitem{p-ttpcc-02}
R.~D.~M. Page, editor.
\newblock {\em Tangled Trees: Phylogeny, Cospeciation, and Coevolution}.
\newblock University of Chicago Press, 2002.

\bibitem{stt-mvuhs-81}
K.~Sugiyama, S.~Tagawa, and M.~Toda.
\newblock Methods for visual understanding of hierarchical system structures.
\newblock {\em {IEEE} Trans. Systems, Man, and Cybernetics},
  11(2):109--125, 1981.

\end{thebibliography}

\newpage
\appendix
\section*{Appendix}
\renewcommand{\textfraction}{0.1}
\begin{figure}[h]
  \centering
  \includegraphics{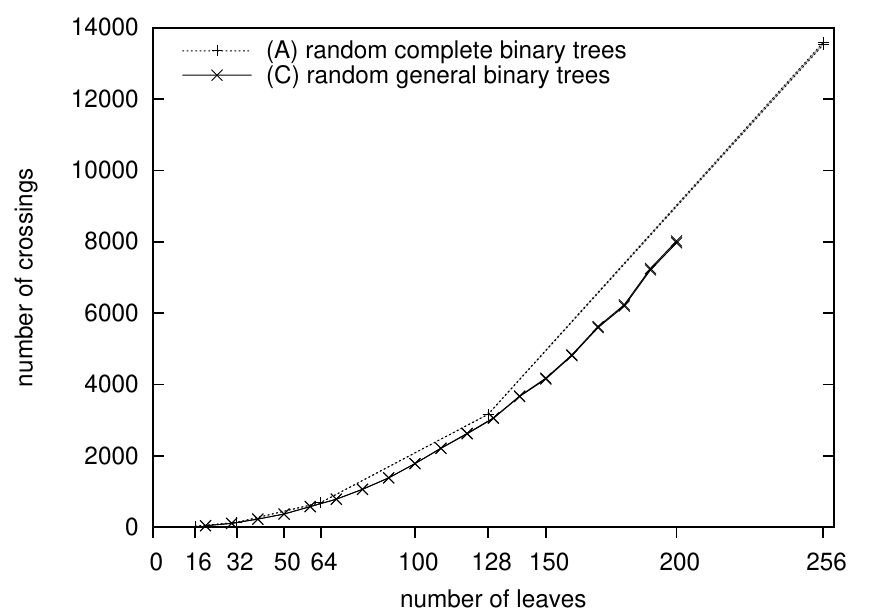}\\
  \includegraphics{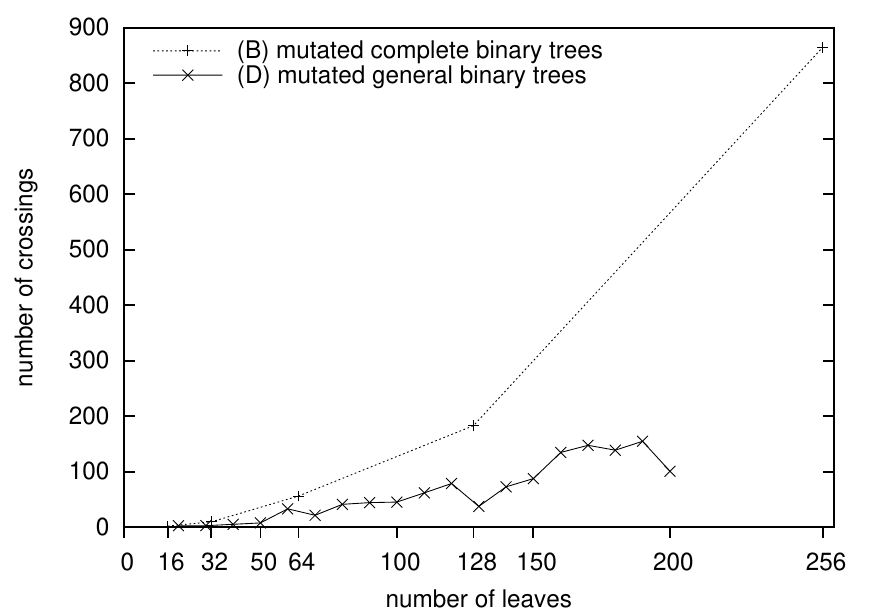}
  \caption{Average crossing numbers of our randomly-generated
    instances. For random (non-mutated) trees in sets~A and~C (top)
    with $n \ge 70$ there is a remaining gap between the best found
    integer solution and the best fractional solution. Both values are
    plotted; however, the gap is relatively small (less than 70 for $n
    \le 256$) and hardly visible.}
  \label{fig:crnumbers-random}
\end{figure}

\end{document}